\documentclass[reprint,showpacs,preprintnumbers,amsmath,amssymb, aps]{revtex4-1}
\usepackage{graphicx,wasysym}% Include figure files
\usepackage{dcolumn}% Align table columns on decimal point
\usepackage{bm}% bold math
\usepackage{natbib}
\usepackage{tabulary}
\newcommand{\kB}{k_{\mathrm{B}}}
\newcommand{\ts}{\tau_{\mathrm{s}}}
\newcommand{\tb}{\tau_{\mathrm{b}}}
\begin{document}

\preprint{1}

\title{Solvation at Aqueous Metal Electrodes}% Force line breaks with \\

\author{David T. Limmer}
\affiliation{Department of Chemistry, University of California, Berkeley, California 94720, United States}
\author{Adam P. Willard}
\altaffiliation{Current address: Department of Chemistry, University of Texas, Austin, Texas 78712, United States}
\affiliation{Department of Chemistry, University of California, Berkeley, California 94720, United States}
\author{Paul A. Madden}
\affiliation{Department of Materials, University of Oxford, Oxford OX1 3PH, United Kingdom}
\author{David Chandler}
\email{chandler@berkeley.edu}
\affiliation{Department of Chemistry, University of California, Berkeley, California 94720, United States}
\date{\today}
\begin{abstract}
We present a study of the solvation properties of model aqueous electrode interfaces. The exposed electrodes we study strongly bind water and have closed packed crystalline surfaces, which template an ordered water adlayer adjacent to the interface. We find that these ordered water structures facilitate collective responses in the presence of solutes that are correlated over large lengthscales and across long timescales. Specifically, we show that the liquid water adjacent to the ordered adlayers forms a soft, liquid-vapor-like interface with concomitant manifestations of hydrophobicity. Temporal defects in the adlayer configurations create a dynamic heterogeneity in the degree to which different regions of the interface attract hydrophobic species.  The structure and heterogeneous dynamics of the adlayer defects depend upon the geometry of the underlying ordered metal surface. For both 100 and 111 surfaces, the dynamical heterogeneity relaxes on times longer than nanoseconds.  Along with analyzing time scales associated with these effects, we highlight implications for electrolysis and the particular catalytic efficiency of platinum.
\end{abstract}

\maketitle
Extended metal interfaces play a fundamental role in aqueous electrochemistry, a field of principle importance in the advancement of renewable, clean energy sources \cite{Somorjai:2010p8235,Nrskov:2009p7956}. In many processes that occur at metal interfaces, such as electrolysis, corrosion and electrocatalysis, water is ubiquitous, often acting as both solvent and reactant\cite{Vayenas:2002p8890}. While many studies exist detailing the behavior of water across small length and timescales, \cite{Schiros:2010p8060,Carrasco:2011p8145,Siepmann:1995p4868,Michaelides:2001p8067} and at low temperatures \cite{Feibelman:2011p6303,Ogasawara:2002p8059}, at present there is little understanding of the large lengthscale correlations and emergent behavior of water on metal surfaces, even though such effects are likely to influence function in important ways \cite{Willard:2008p8256,Han:2012p8065,Carrasco:2012p3354,Lucas:2009p7654}. Here, we report classical simulation results for water on an atomistic model electrode where a detailed study of solvation at the surface has been undertaken. In this study we place particular emphasis on long time correlations and large lengthscale phenomena. Specifically we illustrate how an electrode can impose geometrical constraints within the adlayer of water, creating a composite metal-water interface that is hydrophobic on large length scales. We further show how defects within the hydrogen bonding patterns of the adlayer create transient regions of hydrophobic behavior that exist on small length scales and over long times scales. These results offer a microscopic explanation and generalization of previous experimental observations that have inferred surface hydrophobicity of a platinum electrode at low temperatures \cite{Kimmel:2005p8814,VanDerNiet:2008p8068}.

\begin{figure}[b]
\begin{center}
\includegraphics[width=8.5cm]{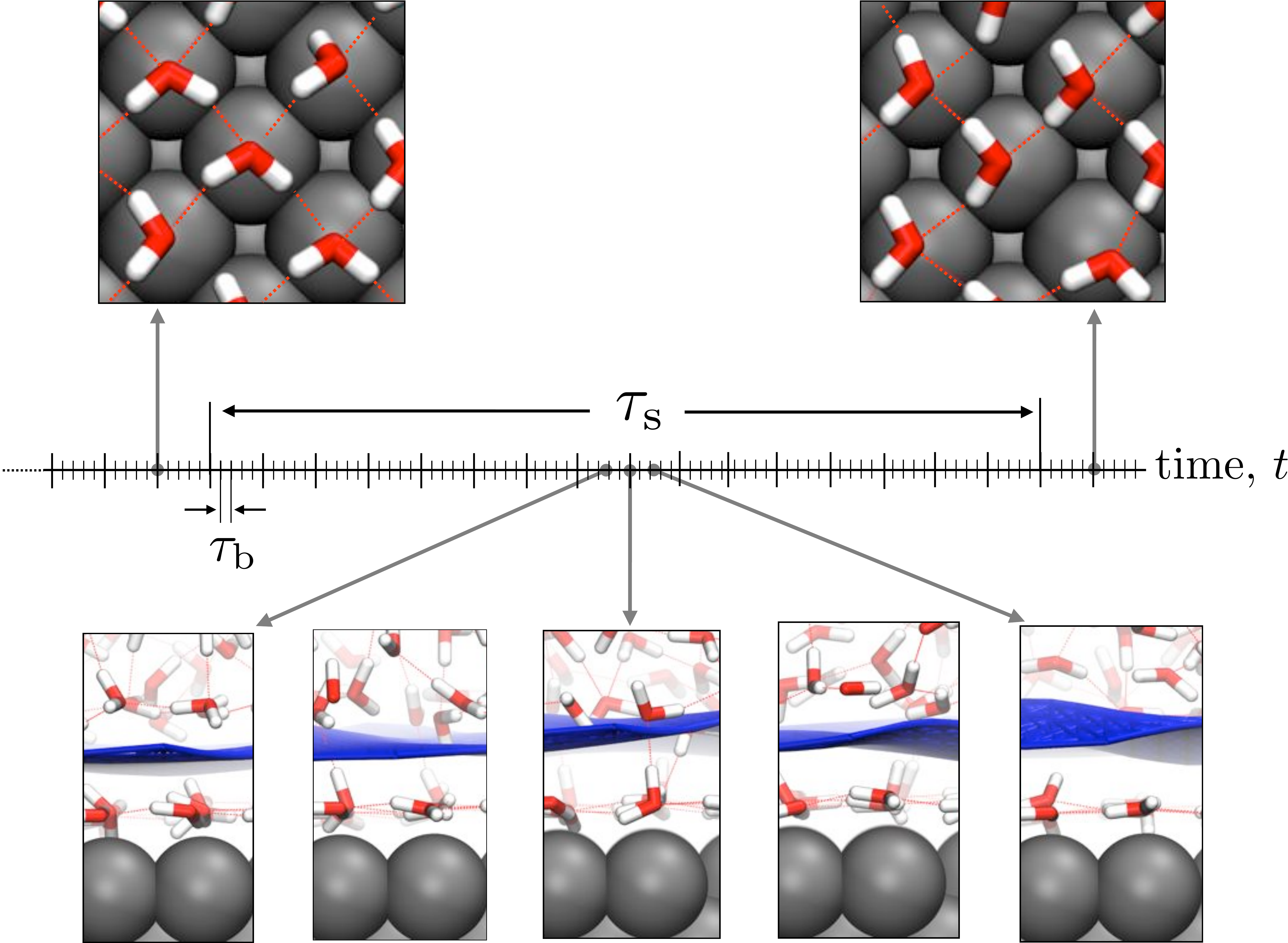}
\caption{Illustration of the separation of timescales between reorganizing surface configurations, which occur on average every $\ts$, and reorganizing the bulk density, which occur on average every $\tb$. Thin tic marks are separated by 20 ps, which is on the order of though larger than timescales for typical density fluctuations. Thick tick marks are separated by 100 ps, which is of the order of the typical relaxation times for relevant interfacial fluctuations. }
\label{Fi:time}
\end{center} 
\end{figure}

To study the electrode interface we use molecular models \cite{Berendsen:1987p8660,Siepmann:1995p4868} that neglect explicit electronic degrees of freedom beyond accounting for electronic polarization of the electrode. Despite its relative simplicity, the model shows reasonable agreement with experiment for the potential of zero charge and capacitance\cite{Willard:2008p8256} of the aqueous platinum interface.  For example, we can estimate the potential of zero charge, $U_\mathrm{pzc}$ relative to the hydrogen electrode following previous work\cite{Cheng:2012p11245} by measuring the contact potential between the electrode and the water bilayer, $\psi\approx-0.9 \mathrm{V}$, and referencing it to the known work function for platinum, $W=5.9\mathrm{V}$ \cite{Cheng:2012p11245} and the absolute hydrogen electrode potential, $U_\mathrm{H_2}=4.4\mathrm{V}$. Using this approach we get $U_\mathrm{pzc}= 0.6\mathrm{V}$ compared to the experimental value of $U_\mathrm{pzc}= 0.4\mathrm{V}$\cite{Jinnouchi:2008p245417}. %For the capacitance, $C$, neglecting  the electronic dipole of the metal can be simply accounted for based on the empirical observation that the induced electronic dipole due to the presence of water does not change much with the specific configuration or surface concentration\cite{Jinnouchi:2008p245417} using a jellium model\cite{Schmickler:1985p1657} yielding $C\approx 40 \mu$F/cm$^2$.} of the aqueous platinum interface \cite{Willard:2008p8256}.

Using this model we can access length and time scales far beyond those currently available to ab inito calculations, and probe collective behavior. We find that the hydrophobicity of the water-electrode interface depends on the amount of passivation of the hydrogen bond network within the adsorbed water layer.  Strong, favorable interactions of the order of half an $e$V between the water and electrode pin the oxygens of the water to the top sites of the crystal lattice creating a spatially ordered arrangement of molecules. We have studied two surfaces of a planar electrode, corresponding to the 100 and 111 faces of an FCC crystal, with lattice spacings corresponding to a platinum lattice. On both the 100 and 111 surfaces, the imposed water structures allow for facile hydrogen bonding within the adlayer and subsequently only a few, fleeting, hydrogen bonds are donated from the adlayer to the surrounding bulk. While the adsorbed oxygens at both surfaces still afford hydrogen bond acceptor sites, the asymmetry associated with lacking donor sites results in an interface that is liquid-vapor like in the sense that  large density fluctuations occur though the collective formation and deformation of an interface\cite{Chandler:2005p640}.

Even though the underlying electrode lattices we study are perfectly ordered, over large length scales the planar geometry of the surface is incommensurate with water's preferred tetrahedral structure. A consequence of this frustration is the presence of an equilibrium number of defects in the hydrogen bond network within the adlayer. These defects facilitate reorganization within the surface, as detailed elsewhere\cite{Willard:preprint}, and the resulting dynamics are heterogeneous and relax on timescales of 1-10's of nanoseconds. This timescale for surface relaxation, which we denote $\ts$, is much larger then typical times for equilibrium density fluctuations, $\ts \gg \tb \approx 5 \mathrm{ps}$. Therefore, while the presence of the surface introduces a static inhomogeneity into the system, the water bound to this surface introduces a dynamic inhomogeneity into the system. The resultant separation of timescales between bulk and surface reorganization is shown in Fig. \ref{Fi:time}. The top panels of Fig.\ref{Fi:time} show snapshots during slow reorganization of the surface water dipoles, while the bottom panels of Fig.\ref{Fi:time} illustrate faster interfacial fluctuations. 

These two features of the electrode interface, the static heterogeneity of the extended interface and the slow dynamics of water at its surface, cause a decoupling of ensemble and dynamic averaging on timescales $t<\ts$. For a given configuration of the adlayer the liquid water in contact with it swiftly equilibrates and for $t\gg\tb$ the properties of the subsequent solvation layers are stationary. Over timescales intermediate between surface and bulk relaxation, we find that the temporal heterogeneity of the hydrogen bond network couples to the solvation properties of the interface, creating transient regions of favorable and unfavorable solvation. These regions of differing solvation interconvert on the timescales of surface relaxation, and only for $t>\ts$, does the solvation at the interface reflect the homogenous symmetry of the fluid above it. 

\begin{figure}[b]
\begin{center}
\includegraphics[width=8.5cm]{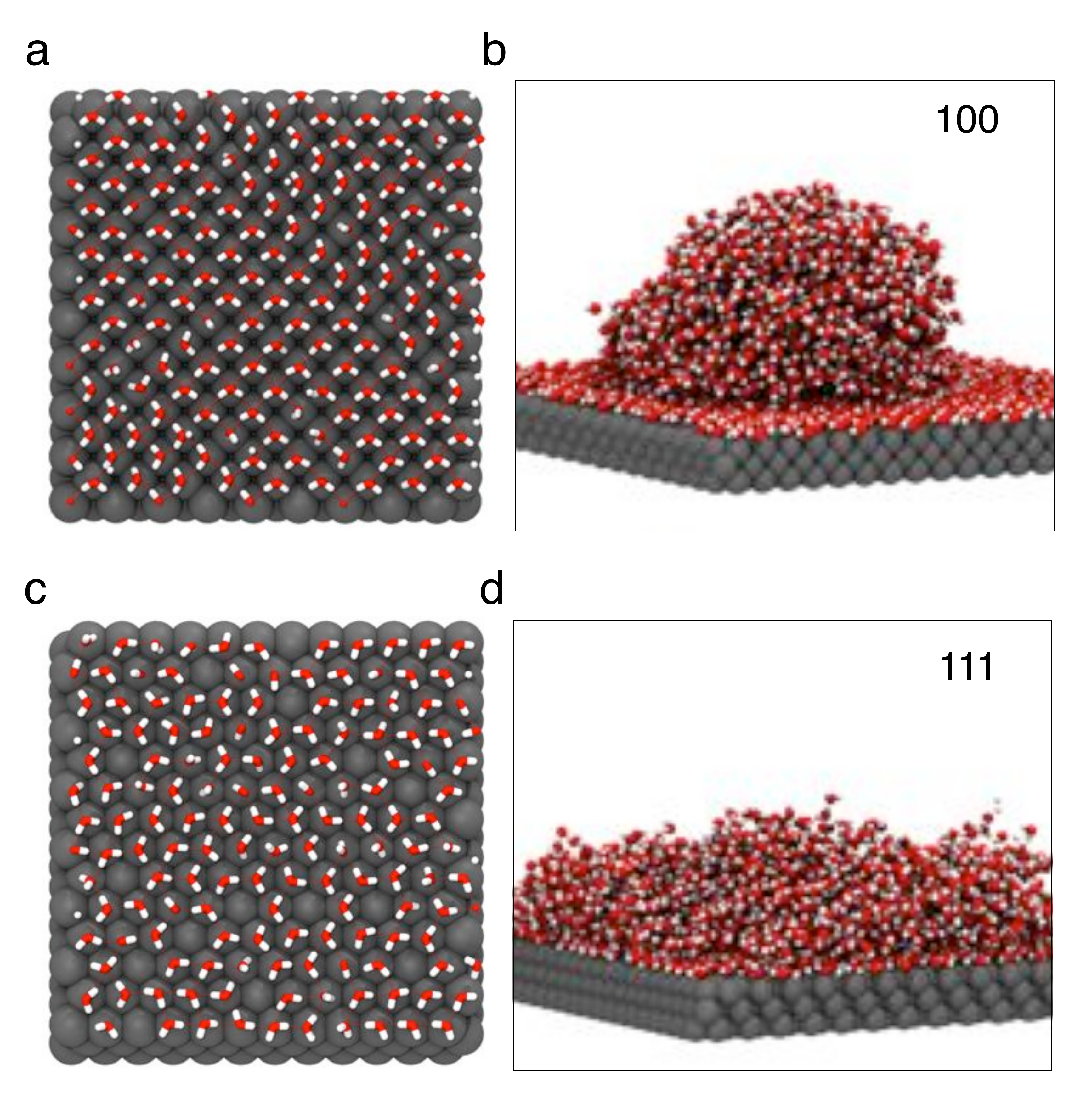}
\caption{Illustration of the 100 and 111 adlayers and their effect on macroscopic solvation. (a) The 100 surface is locally four coordinated and commensurate with favorable hydrogen bonding patterns. Large ordered domains are separated by line defects. (b) The highly ordered domains donate few hydrogen bonds to the subsequent water layers leading those layers to not wet the composite water electrode surface. (c) The 111 surface is locally six coordinated and frustrates hydrogen bonding. Though water is still ordered, vacancies and interstitial defects are common. (d) Hydrogen bond donor sites are more common strongly than on the 100 surface, and subsequently the surface is wet.}
\label{Fi:system}
\end{center} 
\end{figure} 
\begin{figure*}[t]
\begin{center}
 \includegraphics[width=18cm]{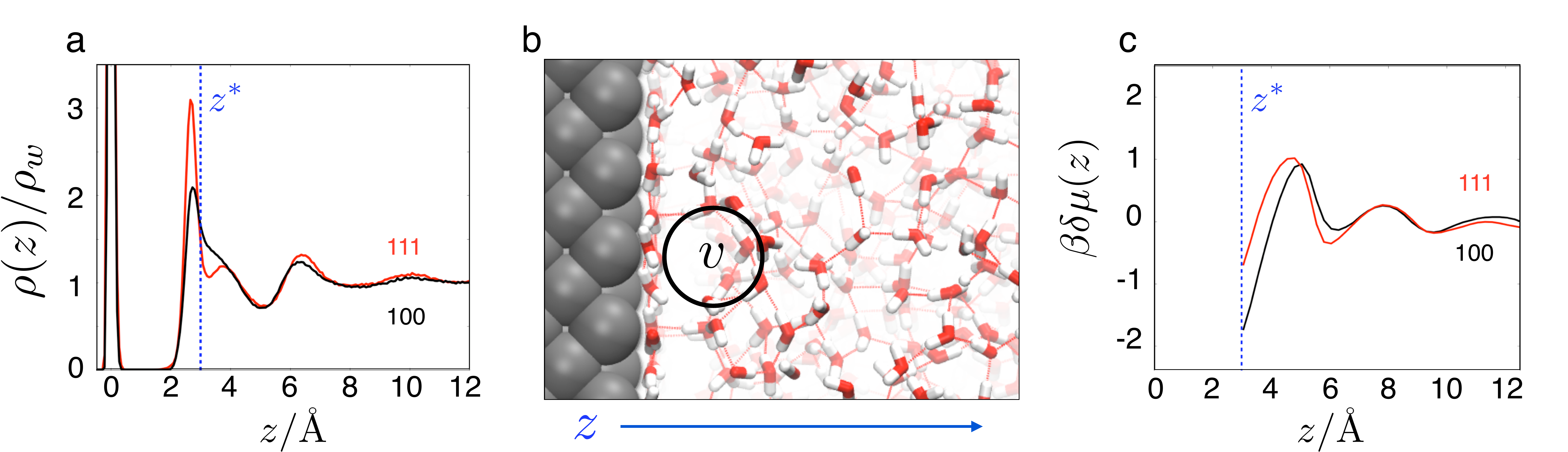}
\caption{Structure and solvation of the composite water-electrode interface. (a) The density profile of water molecules away from both surfaces, divided by the bulk water density, $\rho_w=0.033 \mathrm{A}^{-3}$. The density is characterized by a strongly bound adsorbed layer and solvent layering that extends roughly 10 \AA\, into the bulk. Immediately adjacent to the adsorbed layer is a region of density depletion. (b) For the 100 surface, shown, or a 111 surface, a typical configuration of water molecules shows very little hydrogen bonding (dashed red lines) between the surface and the bulk. The volume $v$ (black circle) is a 3 \AA\, radius sphere. (c) The excess solvation free energy for a 3 \AA \, in radius \, ideal hydrophobe, shown to scale in (b), as a function of distance away from the surfaces. The dashed lines in (a) and (c) define $z^{*}$ as the distance of closest approach of the 3 $\mathrm{\AA}$ sphere to the adlayer.}
\label{Fi:mu_z}
\end{center} 
\end{figure*}

The rest of the paper expands on these findings. First, we illustrate the particular structure of the adlayer and show how the passivation of the hydrogen bond network on the surface creates a liquid-vapor-like interface that easily accommodates small, spherical and large, extended hydrophobes. Next, we show how the strain in the water structure on the surface, coupled with a separation of relaxation times between the surface and bulk, creates temporal regions of spatially heterogeneous solvation potential that decay over nanoseconds. Finally we discuss how the presence of a liquid-vapor like interface is expected to influence electrochemical properties by attracting specific ions such as excess protons necessary for chemical reactions.

\section{Water Electrode Interface} 
The properties of the electrochemical interface are dictated in large part by the first adsorbed water layer. The generic features of this surface are that 1) single water molecules bind strongly and specifically to the top sites of the metal and 2) the crystalline geometries of these top sites are incommensurate with extended hydrogen bonding patterns. Quantum chemical calculations \cite{Carrasco:2011p8145} and experimental photoelectron spectroscopy measurements\cite{Ogasawara:2002p8059} have estimated the single molecule binding energy of water a platinum electrode to be close to 0.4 eV, with a small dependence on surface geometry. The model used here \cite{Siepmann:1995p4868} has been parameterized to recover this strong attraction with binding energies of 0.46 eV and 0.37 eV for the 100 and 111 surfaces respectively. Depending on the specific surface geometry, the thermodynamically stable structure may not be a complete monolayer. This is due to the competition between surface adsorption and hydrogen bond formation within the surface, which is frustrated on surfaces incommensurate with water's preferred hydrogen bonding network which favors four fold coordination\cite{Hodgon:2009p381}.

This interplay of energy scales is reflected in the structural motifs present on the different crystal faces. Figure~\ref{Fi:system} shows characteristic snapshots of the adlayer of water for both surfaces (a and c) as well as their subsequent effect on wetting (b and d). For the 100 surface, metal atoms are locally four fold coordinated and are thus moderately commensurate with a two-dimensional projection of local hydrogen bonding patterns. As a result, the structure of water on the surface is highly ordered with the water's dipole oriented parallel to the surface and approximately all top sites are occupied. At any particular instant, however, line defects exist on the surface separating planes of dipole aligned molecules by 90 degree turns in their orientations. For the 111 surface, metal atoms are locally six fold coordinated and though they also have lattice spacings that are commensurate with a hydrogen bond, the six-fold coordination frustrates preferred bonding patterns. As a result this surface has regions of local hexagonal order, rings are seen in the  structures adopted by monolayers of water absorbed on the 111 surface of  many face-centered cubic metals such as Pt and Pd \cite{Tatarkhanov:2009p8437}, together with an equilibrium concentration of interstitials that occupy empty top sites with dipoles that point away from the surface. This disorder results in an average coverage of about 85\% of all top sites. For both surfaces the lattices are entirely regular, and therefore the heterogeneity in the hydrogen bonding network is temporary. However the imposed order within the adlayer dictates that relaxation occurs over long timescales\cite{Willard:preprint}.  

The presence of extended interfaces in solution, such as the solvated electrode surface, are expected to influence the properties of subsequent solvent layers over distances corresponding to the bulk correlation length. For a liquid near coexistence with its vapor, such as water at ambient conditions, extended inhomogeneities can give rise to a de-wetting transition \cite{Chandler:2005p640}, whose interfaces subsequently have larger correlation lengths. Figure~\ref{Fi:mu_z}(a) illustrates the mean density of water molecules as a functions of the distance away from the electrode surface. Although the structure on the adlayer depends intimately on the electrode geometry, the surrounding water is fairly insensitive to the exposed crystal face. We find for both surface geometries that the density profile for water away from the interface exhibits a sharp peak at the electrode surface, indicative of the adlayer, followed by a region of a density depletion approximately 3 \AA \, thick. Density oscillations decay over 1 nm away from the surface. The region of depleted density demonstrates that the asymmetry between hydrogen bond donors and acceptors at the interface results in an unbalanced attraction, however the effective interaction with the surface is not so weak so as to allow the formation of capillary waves that would destroy the density oscillations seen away from the electrode. 

The density depletion seen immediately adjacent to the adlayer is enough to make solvation of ideal hydrophobes (hard-spheres) favorable at the interface. Figure~\ref{Fi:mu_z} (b) illustrates the size of a 3 \AA \,in radius sphere positioned near the interface. In this representative snapshot it is clear that typical hydrogen bonding patterns near the interface easily accommodate the solvation of such probe volumes. Figure~\ref{Fi:mu_z} (c),  illustrates the excess chemical potential for a hard sphere with a radius of 3\,\AA. We calculate this quantity by monitoring the number fluctuations within a probe volume. Specifically, we calculate the probability of observing $N$ molecules in the probe volume,  $v$, 
\begin{eqnarray}
P_{v}(N) &=& \langle \delta (N_v- N) \rangle \notag \\ 
&=& \lim_{t\rightarrow \infty} \frac{1}{t} \int_0^{t} \,dt' \, \delta (N_v(t')- N) \, ,
\end{eqnarray}
where $\delta(N_v- N)$ is a Kroniker delta function and $\langle \dots \rangle$ denotes equilibrium average. This distribution is related to the excess solvation free energy for an ideal hydrophobe through the relation\cite{Hummer:1996p10041}
\begin{equation}
P_v(0) = e^{-\beta \Delta \mu_v} 
\end{equation}
where $\Delta \mu_v$ is the reversible work to create a cavity of size and shape $v$ and $\beta$ is Boltzman's constant. 

For the systems we consider here, the existence of the planar electrode breaks transitional invariance. In order to accommodate this aspect we denote, $P_{v(\mathbf{r})}(N)$, where $\mathbf{r}$ is the position of the center of the probe volume. This distribution reduces to $P_{v(\mathbf{r})}(N) \rightarrow P_{v}(N)$ when $\mathbf{r}$ is far away from the electrode surface. Correspondingly, we also define  $P_{v(\mathbf{r})}(0) = \exp{[-\beta \Delta \mu_{v(\mathbf{r})}}] $. In other words, solvation free energy in an inhomogeneous system is generally spatially dependent.  

Due to the separation of timescales between surface and bulk relaxation, our system is also dynamically heterogeneous. Therefore, on intermediate timescales, $\tb < < t < \ts $, the solvation free energy carries a time-dependence.  This time dependence is denoted as, 
\begin{eqnarray}
P_{v(\mathbf{r})}(N,t;\mathbf{x}_0) &=& \frac{1}{t} \int_0^{t} \,dt' \, \delta \left [ N_{v(\mathbf{r})}(t';\mathbf{x}_0)- N \right ] \\
&=& e^{-\beta \Delta \mu_{v(\mathbf{r})}(t; \, \mathbf{x}_0)} 
\label{Eq:Pv_t_r}
\end{eqnarray}
where $\mathbf{x}_0$ denotes the initial surface configuration and $t$ is the timescale over which the distribution is averaged. For $t  \ll \ts$, Eq. \ref{Eq:Pv_t_r} simplifies to,
$P_{v(\mathbf{r})}(N, t; \, \mathbf{x}_0 ) \rightarrow P_{v(\mathbf{r})}(N)$. 
\begin{figure}[bh]
\begin{center}
\includegraphics[width=7.cm]{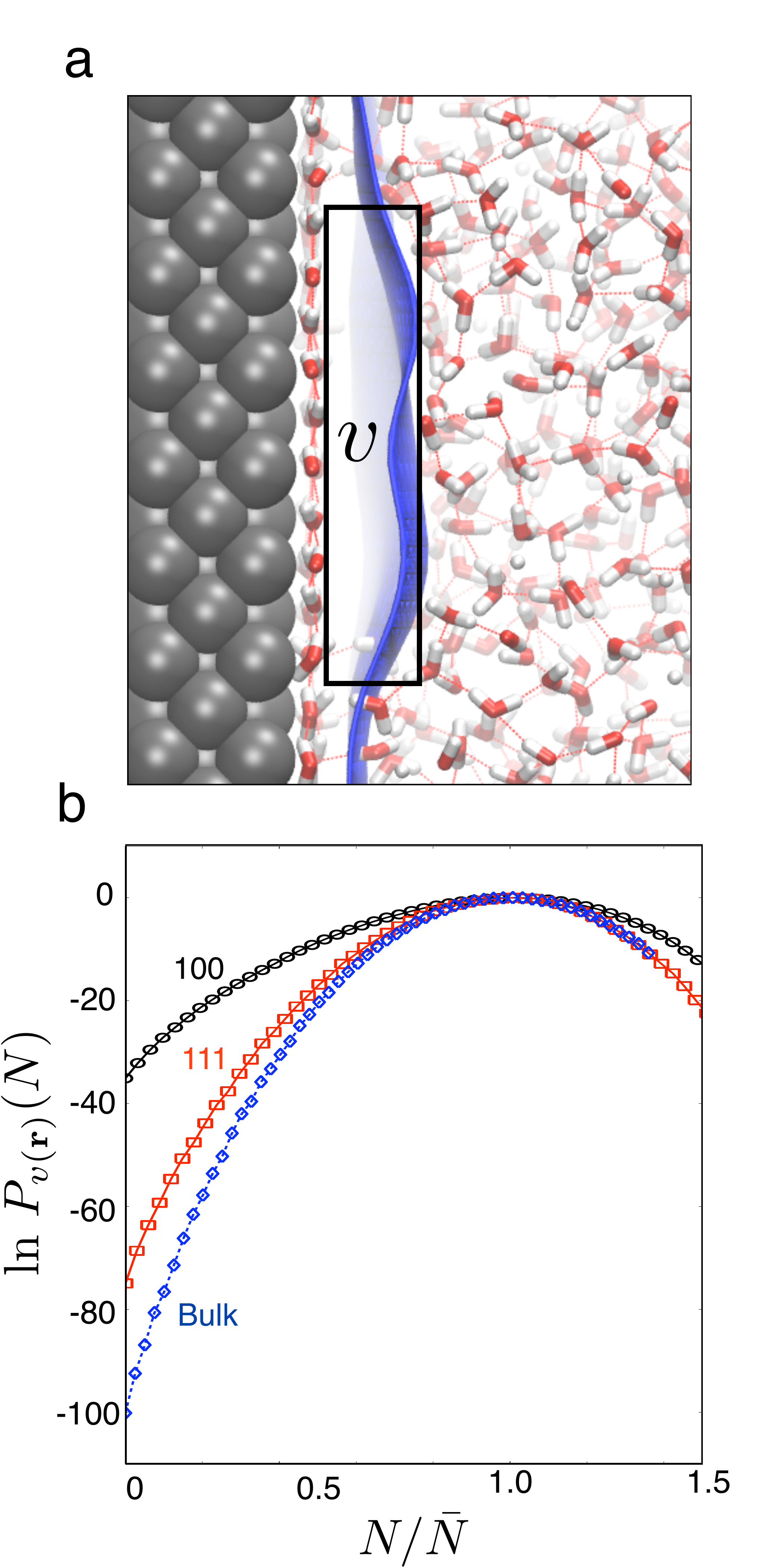}
\caption{Solvation of a large probe volume near the interface. (a) A configuration of water molecules near a 100 surface with the instantaneous liquid interface shown in blue\cite{Willard:2010p1954}. This configuration is typical of one solvating a cuboid of size 20 by 20 by 3 \AA, shown to scale in black. (b) Probability distribution for finding $N$ particles in a probe volume whose outer edge is located at $z^*$, with a mean occupancy $\bar{N}$. $\bar{N}$ = 31,34 and 40 for the 100 surface, 111 surface and bulk, respectively. }
\label{Fi:bigv}
\end{center} 
\end{figure}
Finally, the difference between the value of the solvation free energy located at $\mathbf{r}$, averaged over a time $t$, and its equilibrium bulk value is defined as 
\begin{equation}
\delta \mu_v(\mathbf{r},t; \, \mathbf{x}_0) = \Delta \mu_{v(\mathbf{r})}(t; \, \mathbf{x}_0) - \Delta \mu_v \, .
\end{equation}
For $t\rightarrow \infty, \delta \mu_v(\mathbf{r},t; \, \mathbf{x}_0) \rightarrow \delta \mu_v(\mathbf{r})$ and for $\mathbf{r}$ far away from the surface $\delta \mu_v(\mathbf{r}) \rightarrow 0$. 
At long times, and averaged over the plane of the surface, the solvation free energy has a minimum at the distance of closest approach to the composite water-electrode surface, $z^{*}$, indicated by a dashed line in Fig.~\ref{Fi:mu_z}. The negative solvation free energy implies that while the bare electrode surface attracts water, the composite water-electrode surface preferentially attracts oil and can be considered hydrophobic. While both electrode geometries exhibit enhanced hydrophobic solubility, the 100 surface is more hydrophobic as measured by its excess solvation free energy at $z^{*}$, $\beta \delta \mu(z^{*}) \approx -2.0$, compared to the 111 surface, $\beta \delta \mu(z^{*}) \approx -0.7$ for the 3 \AA\, sphere.

\section{Large Lengthscale Solvation} Using the method of indirect umbrella sampling (INDUS) \cite{Patel:2011p8761} we are able to compute stationary distribution functions, $P_{\mathrm{v}(\mathbf{r})}(N)$, for extremely rare fluctuations. By studying the tails of these distributions we can determine to what extent interface formation, as opposed to Gaussian density fluctuations, are important.

The specific dimensions of the probe volume are chosen to focus on interfacial fluctuations.  In particular, the probe volume is thin enough, 3 \AA, to include molecules that can be part of a liquid interface while not also containing molecules that are part of the bulk; and it is wide enough, $20 \times 20 \mathrm{\AA}^2$, to capture nano-scale fluctuations intrinsic to the soft liquid interface. Figure~\ref{Fi:bigv} (a) illustrates a configuration of water and the instantaneous liquid interface\cite{Willard:2010p1954} solvating the cuboid sub-volume, highlighting how solvation at the surface occurs by deforming a soft interface. The signature of this mode is illustrated in Fig.\ref{Fi:bigv} (b) where highly correlated behavior, interface formation in this case, is apparent by the appearance of a exponential tail for small $N$ for both surfaces relative to the bulk.  These distributions along with Eqs. 2 and 4 allow us to calculate the excess solvation free energy. We find that this is negative at both surfaces, however solvation for this large probe volume at the 100 surface is more favorable by $40 \,\kB T$ compared to the 111 surface.  While both surfaces afford hydrogen bond acceptor sites, only the 111 surface has a nonzero number of fleeting hydrogen bond donors. Using a standard criteria for defining a hydrogen bond\cite{Luzar:1996p928}, we calculate an average hydrogen bond donor density on the surface to be approximately 1.0/nm$^2$ for the 111 surface and 0.0/nm$^2$ for the 100 surface.   This means that in the large probe volume, there is on average 4 hydrogen bonds donated to the bulk. These few hydrogen bonds, produce the $40 \,\kB T$ change in solvation, and it is expected that the addition of further hydrogen bond defects would make the surface hydrophilic.

\begin{figure}[bth]
\begin{center}
\includegraphics[width=7cm]{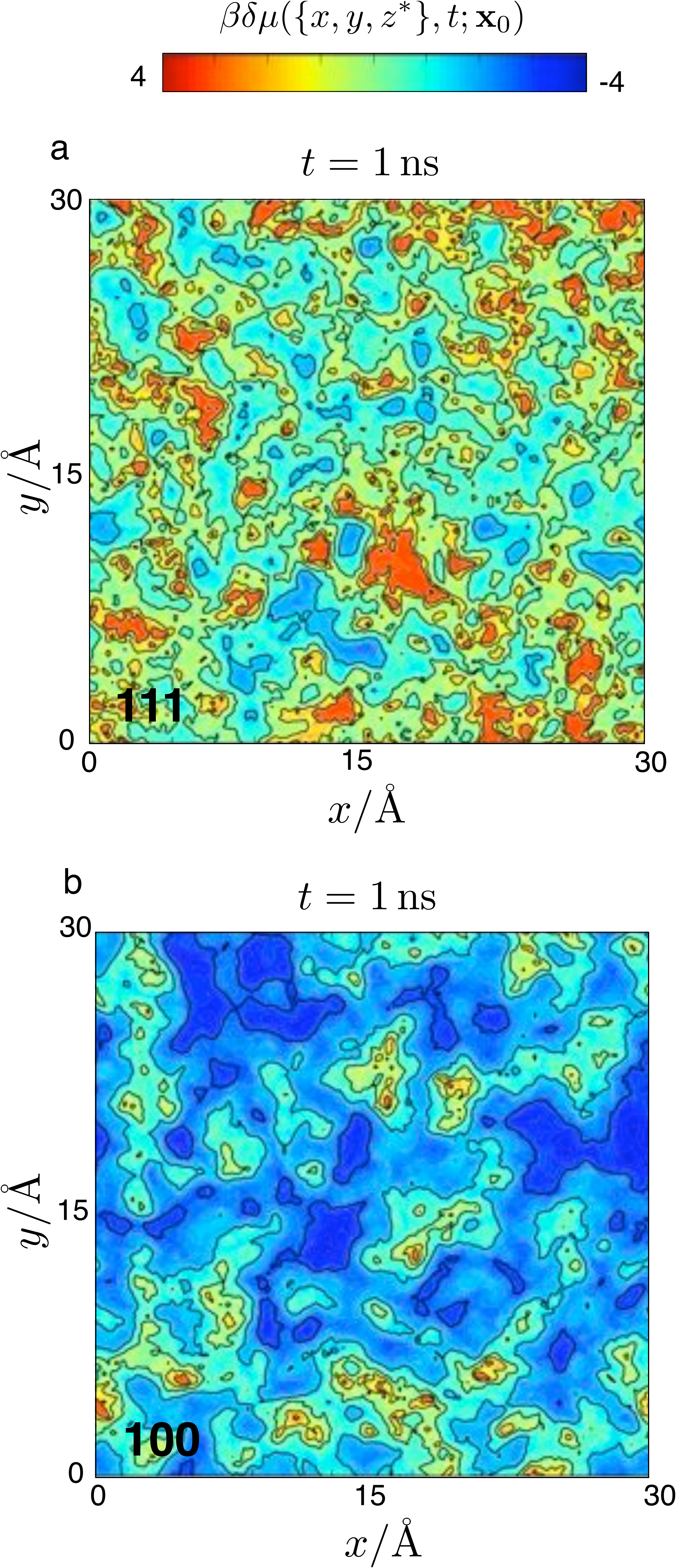}
\caption{Heterogeneous solvation at the electrode surface. (a,b) The excess free energy for a 3 \AA \, ideal hydrophobe is heterogeneous at both 111 and 100 surfaces. Regions of favorable and unfavorble solvation have been determined by averaging over 1 ns from an initial surface configuration, $\mathbf{x}_0$.}
\label{Fi:surface}
\end{center} 
\end{figure}
This microscopic measure of solvation explains the different wetting behaviors observed for these surfaces in Fig. \ref{Fi:system}. Large lengthscale solvation at hydrophobic surfaces is dominated by deforming existing interfaces, the excess chemical potential is expected to be well approximated by
\begin{equation}
\beta \delta \mu(z^{*}) = -A \gamma_\mathrm{LV}(1 - \cos \theta )
\end{equation}

\begin{figure*}[tf]
\begin{center}
\includegraphics[width=16cm]{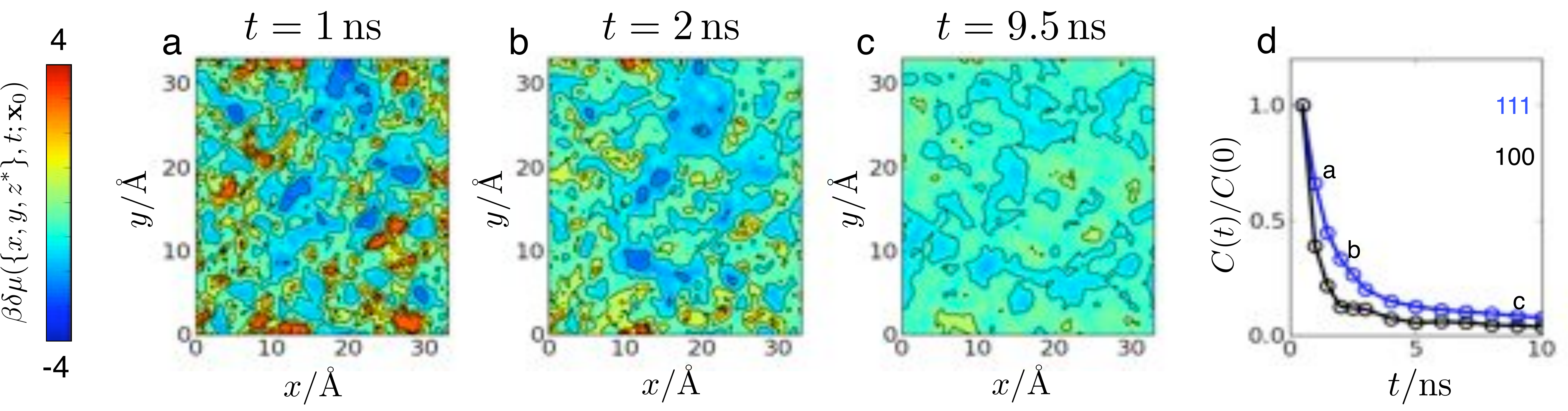}
\caption{Evolution of the solvation free energy as the surface reorganizes from an initial configuration, $\mathbf{x}_0$. (a-c) Maps of the excess free energy averaged over different observation times for the 111 surface. Contour plots are reported using the same scale shown in the colorbar on the left. (d) The time correlation function for solvation heterogeneities for each surface. Points (a), (b) and (c) coincides with times for panels (a), (b) and (c).}
\label{Fi:evolution}
\end{center} 
\end{figure*}
where $A$ is the cross-sectional area of a large probe volume, $\gamma_\mathrm{LV}$ is the liquid-vapor surface tension, and $\theta$ is the water droplet contact angle on the surface. The more favorable solvation at the surface of the 100 surface is expected to result in a contact angle $\theta \approx  90^{\mathrm{o}}$, whereas the subtly favorable solvation at the 111 surface is expected to result in a contact angle of $\theta \approx 40^{\mathrm{o}}$. These are in qualitative agreement with the configurations shown in Fig.~\ref{Fi:system}. The connection with macroscopic wetting behavior presented here is also in agreement of previous observations of hydrophobicity of the adlayer for the platinum 111 surface\cite{Kimmel:2005p8814}, and suggests that studies probing the platinum 100 surface would find an even more pronounced effect.  

\section{Heterogeneous Surface Solvation}
The data presented in Fig. \ref{Fi:mu_z} illustrates equilibrium values of the excess solvation free energy averaged over the plane parallel to the surface, computed by averaging over long molecular dynamics trajectories, of roughly 10 ns. The presence of a region of strong water density depletion induced by the local structure of the water adlayer implies that only on timescales much longer than the correlation time for typical bulk density fluctuations, $\tb \approx 5 \mathrm{ps}$ \cite{English:2011p2938}, will solvation within this plane should be homogeneous, i.e. $\beta \delta \mu(x,y) \approx \mathrm{const}$. However, the ordering within this surface adlayer makes reorganization difficult, and as a result the timescale associated with de-correlating surface configurations, $\ts \approx 1-10 \mathrm{ns}$, is long (see Appendix A). For intermediate times, $\tb \ll  t < \ts$, this long time surface relaxation couples to the solvation in interesting ways. In particular, we find that for averages computed over this time, $1\, \mathrm{ns} < t < 10 \, \mathrm{ns}$, the solvation calculated within the surface is heterogeneous. 

Figure~\ref{Fi:surface} depicts the solvation free energy, $\beta \delta \mu(\{x,y,z^{*}\},t; \mathbf{x}_0)$ at both surface geometries for a 3 \AA \, sphere, as a function of position in the $x,y$ plane and observation time, for a given surface configuration. This method of spatially resolving the local hydrophobicity is a time dependent extension of previous work by others on protein surfaces, which have static heterogeneity\cite{Acharya:2010p10048}. As shown in Fig.~\ref{Fi:surface} the structure of the solvation on this surface reflects neither the underlying lattice symmetry nor the homogeneous symmetry of the above liquid, but rather the coupling between hydrogen bonding defect structures of the bound water adlayer and the above solvent layers.  We can quantify the surface heterogeneity by calculating a time-dependent variance 
\begin{equation}
C(t) =\langle \left  [ \,  \delta \mu(\{x,y,z^{*}\},t; \mathbf{x}_0) -  \delta \mu(z^{*})  \right ]^2 \rangle 
\end{equation}
where $\delta \mu(z^{*})$ is an average excess solvation free energy at $z^*$ and as before $\langle \dots \rangle$ denote averages over realizations of initial surface conditions. We find for times satisfying $t \gg \tb$, the spatial average over the surface is equal to the long time average. Using this measure we find that for all times the solvation on the 111 is more heterogeneous than on the 100 surface, owing to the larger domain sizes seen on the ordered bound layer in the 100 surface.  These domains sizes are on average approximately $1 \mathrm{\AA}^2$ for the 111 surface and  $3 \mathrm{\AA}^2$ for the 100 surface, as obtained by coarse-graining Fig.~\ref{Fi:surface}(a,b) over 1 $k_\mathrm{B}T$. Figure \ref{Fi:evolution} (a-c) show $\beta \delta \mu(\{x,y,z^{*}\},t;  \mathbf{x}_0)$ as $t$ is increased for the 111 surface. Generically, reorganization on the surface occurs as $t$ is increased, and the amount of heterogeneity is reduced. Similar behavior is found for the 100 surface. In order to quantify the timescales for relaxing this heterogeneity we measure the decay of $C(t)/C(0)$, where the argument of the denominator is taken at the smallest $t$ where  $\tb \ll  t \approx 0.5 \mathrm{ns}$. Figure \ref{Fi:evolution} (d) plots $C(t)/C(0)$ for both surfaces and illustrates that the time to reach a uniform solvation potential at the surface is on the order of 10 ns. This time is on the order of many of the slow processes the occur on the electrode surface such as the mean dipole correlation time 1-10 ns (see Appendix A). 
 
\section{Implications for Electrolysis}
The results illustrated here have implications for catalysis, and electrolysis especially. In particular, we have shown that the hydrophobic surface that is formed from a passivated adlayer of water is accompanied by the existence of a liquid-vapor-like interface separating the adlayer from the bulk liquid. It has been previously demonstrated that excess protons\cite{Petersen:2005p7976,Buch01052007}, as well as some anions\cite{Jungwirth:2002p8414}, preferentially adsorb to a liquid-vapor interface. It is expected, therefore, that the existence of this type of interface near the electrode surface itself acts catalytically by simply enhancing the local concentration of reactive species-- protons -- relative to the bulk.  

\begin{figure}[t]
\begin{center}
\includegraphics[width=7.cm]{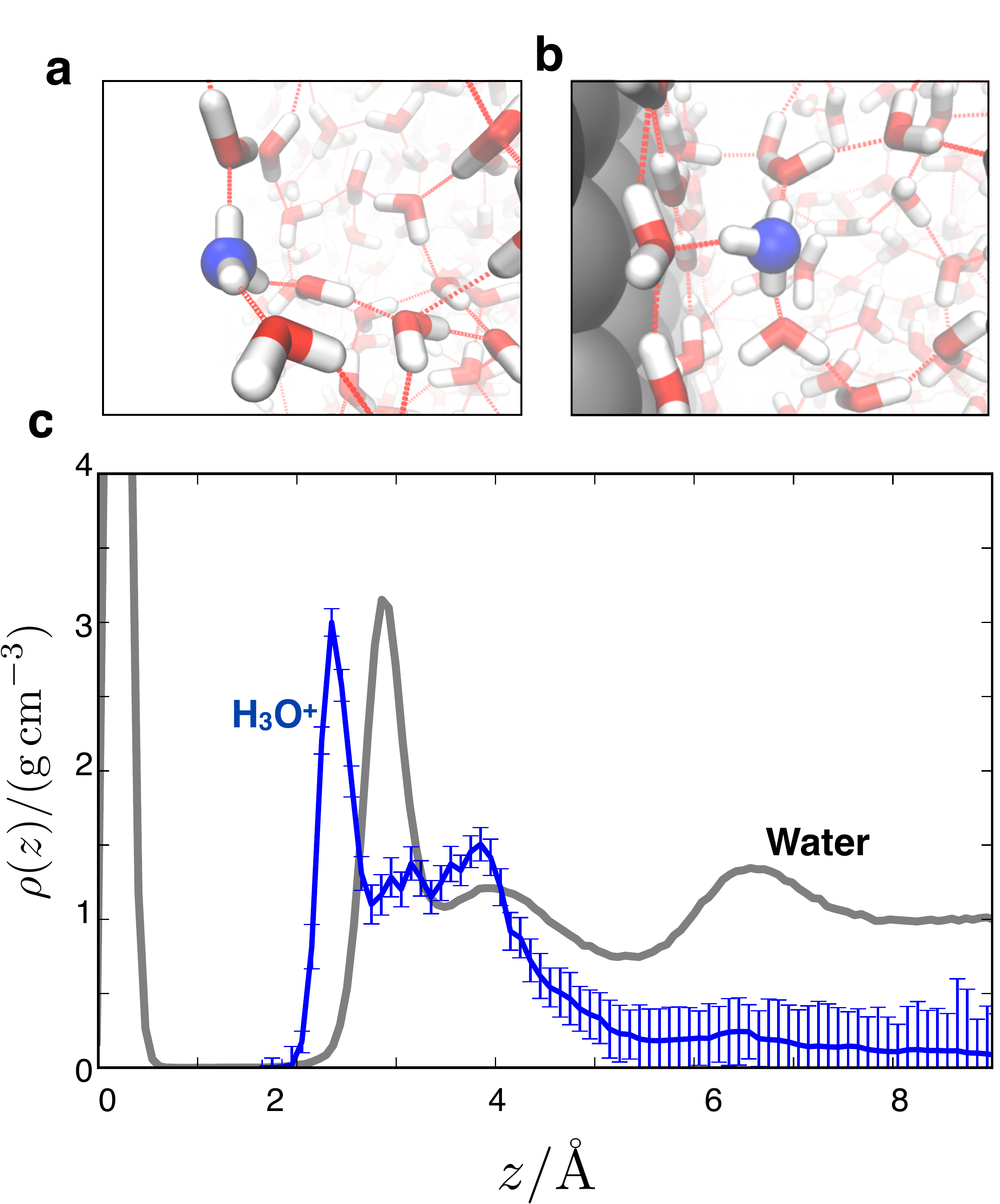}
\caption{Structure and density enhancement of hydronium ions at the liquid-vapor and electrode interface. (a) A characteristic snapshot of a $\mathrm{H}_3\mathrm{O}^+$ at the liquid vapor interface. (b) A characteristic snapshot of a $\mathrm{H}_3\mathrm{O}^+$ at the 111 interface. (c) Density distributions for the water (grey) and hydronium (blue), with a reference bulk density of the hydronium, $\rho_\mathrm{ref}=0.1$g/cm$^3$ chosen for scale.}
\label{Fi:h3oplus}
\end{center} 
\end{figure}

As one simple means of testing the assertion of a proton enhancement at the interface, we have calculated the density profile for a fixed point charge model of a hydronium\cite{Vacha:2008p4975} cation using umbrella sampling. We choose this model as it has been shown previously that excess protons at the liquid vapor interface preferentially adopt a hydronium geometry over other forms, such as the Zundel cation\cite{Petersen:2005p7976,Buch01052007}. The local solvation structure of a hydronium in bulk is characterized by  donation of a hydrogen bond by each of its three hydrogens, but its inability to accept any hydrogen bonds at the oxygen position due to the localized positive charge on the molecule. It has been demonstrated previously that this structure is conserved at the liquid vapor interface, and has been proposed as the reason for experimental observations of proton adsorption at the interface. A characteristic snap shot of such an expected configuration  is shown in Fig.~\ref{Fi:h3oplus} (a). This snapshot is obtained from our simulations, in which find a mild enhancement of $ \rho(0) \approx 1.5 \rho_\mathrm{bulk}$ at the liquid-vapor interface, where the zero refers to the position of the Gibbs dividing surface. 

At the water-electrode interface the adlayer is composed almost entirely of hydrogen bond acceptor sites and thus it is expected that density of hydronium ions will be even further enhanced by their ability to donate hydrogen bonds into the adlayer. A characteristic snap shot of this type of configuration at the 111 surface is shown in Fig. \ref{Fi:h3oplus} (b). Figure \ref{Fi:h3oplus} (c) confirms a much larger density enhancement of hydronium ions at the interface relative to the liquid-vapor case. This density distribution is calculated from, 
\begin{equation}
\rho_{\mathrm{H}_3\mathrm{O}^+} (z)= \rho_\mathrm{ref} \, e^{-\beta F(z)} \, ,
\end{equation}
where $\rho_\mathrm{ref}$ is the density of the hydronium in the bulk and $F(z)$ is the potential of mean force for moving the center of mass of the hydronium along the $z$ direction. The enhancement found from this simplistic calculation is $\rho(z^{*}) \approx 10 \rho_\mathrm{bulk}$, much larger than the enhancement found at the liquid vapor interface. We note that while important aspects of proton delocalization and polarizability are neglected in this calculation, each affect has been shown previously to further enhance interfacial adsorption. Thus while this calculation is meant only as illustrative of our findings, and not expected to be taken quantitatively, we nevertheless suspect this behavior to be conserved in more detailed models.  

In the specific case of  hydrogen evolution at a platinum electrode, it is generally assumed that the reaction proceeds through two steps: the Volmer step, where a proton is transferred form the bulk and discharged at the metal surface, followed by the Tafel step where two adsorbed protons combine to form hydrogen and desorb from the surface\cite{Holger:2010p724}. The latter is considered to be the rate determining step. The enhancement of the proton concentration at the interface is consistent with platinum's ability to easily transfer and accumulate protons on and near the surface, while the long time relaxation on the surface detailed here and elsewhere\cite{Willard:preprint} undoubtably makes diffusion of adsorbents slow.  While these results are consistent with mechanistic assertions for hydrogen evolution, further work must be done to explore the full implications of the effects illustrated here on catalysis. 
 
\section{Methods}
The system simulated consists of a slab of water in contact with an electrode on one side and with a free interface on the other side with a vacuum layer of 40 \AA. The electrode consists of three layers of atoms, totaling nearly 500 particles, held fixed in an FCC lattice with spacing of 3.92 \AA \, and with either the 100 or 111 facet exposed to the solution.  A slab of water nearly 40 \AA\, thick was placed in contact with the electrode, and the dynamics of the nearly 1800 molecules are propagated using a Nose Hover integrator \cite{Martyna:1994p4409}], with SHAKE imposing bond and angle constraints for the water as implemented in LAMMPS \cite{Plimpton:1995p3851}. All simulations were run at 298 K. Periodic boundary conditions are employed in the plane parallel to the surface, and slab corrections to ewald summations are applied.  Interactions between the water molecules are computed from the SPC/E potential \cite{Berendsen:1987p8660}. The water electrode potential is modeled following Siepmann and Sprik \cite{Siepmann:1995p4868} where the platinum water interaction is a sum of two and three body terms, parameterized to get the correct value of the adsorption energy and ground state geometry as determined by quantum chemical calculations. Additionally, to model the polarizable metal surface each electrode atom carries a Gaussian charge of fixed width but variable amplitude, which is updated at each timestep by minimizing the energy of the slab subject to a constraint of equal potential across the conductor.  A more thorough description of the electrode model can be found in \cite{Siepmann:1995p4868,Willard:2008p8256}. 

\begin{figure}[b]
\begin{center}
\includegraphics[width=8.5cm]{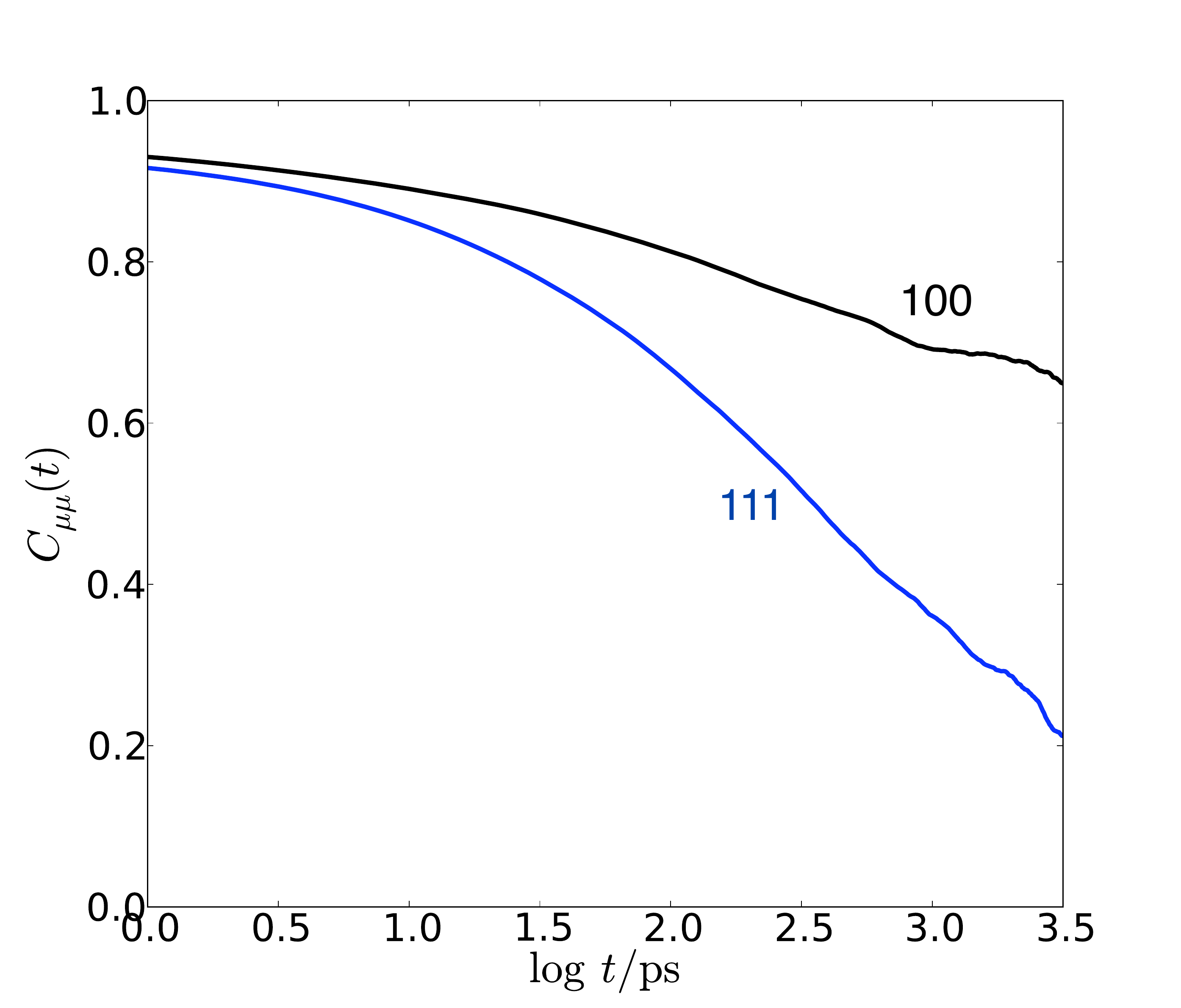}
\caption{The correlation function for dipole orientations for waters on the 100 and 111 surfaces. }
\label{Fi:relax}
\end{center} 
\end{figure}

The long relaxation times in the system relative to the timescales accessible by our simulations make ensuring equilibration on the surface difficult. In order to check the sensitivity of our results to their initial conditions we have prepared an ensemble of 40 independent surfaces, for both the 100 and 111 crystal faces. Of the surfaces, 20 were produced through quenching the system at a rate of 10 K/ns from a system equilibrated at $T=400$, and 20 from a process of vapor deposition where water molecules are exposed to the surface at a rate of 2 molecules/ 1 ns nm$^2$, which corresponds closely to the equilibrium desorption rate. We have redone all of the calculations presented in the main text over this extended surface ensemble and found results that were indistinguishable from those presented above.

The calculations of the excess hydronium ion were accomplish using umbrella sampling along the z coordinate. For charge neutrality a small anion was placed in the water slab but kept at distances greater than 15 \AA \, from the hydronium.

\appendix

\section{Surface Relaxation}
Relaxation on the electrode surface is slow, owing to the large ordered domains patterned by the underlying lattice symmetry. In order to quantify the timescales associated with surface reorganization we calculate a water's dipole time correlation function given it starts and ends adsorbed to the surface, 
\begin{equation}
C_{\mu,\mu} (t) = \langle \vec{\mu}_i(t) \cdot \vec{\mu}_i(0) h[i(t)]h[i(0)]\rangle \, ,
\end{equation}
where $\vec{\mu}_i$ is the dipole vector for the $i$ water molecule and $h_i$ is an indicator function which is equal to 1 if the center of mass of water molecule $i$ is within 3 \AA \, of a electrode atom and is 0 otherwise. 

Figure \ref{Fi:relax} reports on this function for both electrode geometries. Both have a characteristic time constants that decay over timescales longer than 1 ns. The long timescales associated with this relaxation make ensuring equilibration of the simulations difficult. In order to address this we have confirmed that all of our results are independent of the preparation of the initial condition. See methods for details. 

%\bibliography{ref3}
% Produces the bibliography via BibTeX.
\end{document}